\documentclass[aps,prl,twocolumn,superscriptaddress,longbibliography]{revtex4-2}
\usepackage{graphicx}
\usepackage{latexsym}
\usepackage{amssymb}
\usepackage{amsmath}
\usepackage{amsfonts}
\usepackage{upgreek}
\usepackage{float}
\usepackage{bm}
\usepackage{multirow}
\usepackage{color}
\usepackage[T1]{fontenc}
\usepackage{libertine}
\usepackage{hyperref}
\usepackage{subfigure}
\usepackage{xcolor}

\hypersetup{
colorlinks = true,
linkcolor = [rgb]{0.70,0.13,0.13},
citecolor = [rgb]{0.13,0.55,0.13},
urlcolor  = [rgb]{0.25, 0.41, 0.88}}
\newcommand{\ket}[1]{|#1\rangle}
\newcommand{\bra}[1]{\langle#1|}

\begin{document}

\title{Aufbau Principle for Non-Hermitian Systems}

\author{Gaoyong Sun}
\thanks{Corresponding author: gysun@nuaa.edu.cn}
\affiliation{College of Physics, Nanjing University of Aeronautics and Astronautics, Nanjing, 211106, China}
\affiliation{Key Laboratory of Aerospace Information Materials and Physics (NUAA), MIIT, Nanjing 211106, China}
\author{Su-Peng Kou}
\affiliation{Center for Advanced Quantum Studies, Department of Physics, Beijing Normal University, Beijing 100875, China}

\begin{abstract}
We develop a generalized Aufbau principle for non-Hermitian systems that allows for building up the configurations of indistinguishable particles.
The Aufbau rule of non-Hermitian systems is unexpectedly shown to be identical to that developed in Hermitian systems when the real parts of the complex energy levels are considered.
We derive full many-body energy spectra of the fermionic and bosonic Hatano-Nelson models as examples by filling single-particle energy levels in the momentum space.
For open boundary conditions, we show that many-body non-Hermitian skin effects persist in all many-body eigenstates for both fermions and bosons.
Furthermore, we find surprisingly that the ground state of bosons is an anomalous Bose-Einstein condensation with all of the particles simultaneously localizing in both the real and momentum space beyond the Heisenberg uncertainty principle.
For periodic boundary conditions, we show that hard-core bosons cannot be mapped to fermions.
This work establishes a general framework for understanding the many-body physics of non-Hermitian systems, revealing rich unique non-Hermitian many-body physics.

\end{abstract}

\maketitle

{\it Introduction.-} Non-Hermitian systems that have recently attracted great interest are shown to exhibit rich unique phenomena with no counterparts in Hermitian systems \cite{bergholtz2021exceptional,ashida2021non}.
For example, it is argued that non-Hermitian Hamiltonians can give rise to non-Hermitian skin effects \cite{lee2016anomalous,yao2018edge,kunst2018biorthogonal,xiong2018does,
gong2018topological,alvarez2018non,yokomizo2019non,okuma2020topological,zhang2020correspondence,yang2020non,
wang2020defective,jiang2020topological,weidemann2020topological,xiao2020non,borgnia2020non}, 
exceptional points and the bulk Fermi arcs \cite{heiss2012physics,kozii2017non,hodaei2017enhanced,zhou2018observation,miri2019exceptional,park2019observation,yang2019non,
ozdemir2019parity,dora2019kibble,jin2020hybrid,xiao2021observation,chen2022asymmetric}, and non-Hermitian edge bursts \cite{wang2021quantum,xue2022non}.
The non-Bloch band theory is subsequently developed to explain these non-Hermitian phenomena in the single-particle picture \cite{yao2018edge,yokomizo2019non}.
One of the following goals is to explore many-body physics and theories of non-Hermitian systems.
Rich unique many-body physics have been explored beyond Hermitian systems, 
including phase transitions without gap closing \cite{matsumoto2020continuous,yang2022hidden}, many-body skin effects \cite{zhang2020skin,mu2020emergent,lee2021many,kawabata2022many,zhang2022symmetry,alsallom2022fate,shen2022non}, 
many-body edge bursts \cite{hu2023many},
and entanglement phase transitions \cite{kawabata2023entanglement,feng2023absence}.

So far, the approaches studying non-Hermitian many-body systems are mainly limited to numerical simulations \cite{yang2022hidden,zhang2020skin,mu2020emergent,lee2021many,kawabata2022many,zhang2022symmetry,alsallom2022fate,hu2023many,kawabata2023entanglement,feng2023absence,lee2020many,sun2022biorthogonal,tang2022dynamical,hyart2022non,guo2022variational,guo2023composite,shen2023construction,shen2022non,shen2023proposal}, 
the perturbation theory \cite{sun2022biorthogonal,zhang2022symmetry}, and the Green function \cite{hu2023many,hyart2022non}.
The complete understanding of the many-body physics of non-Hermitian systems remains elusive although there are studies based on exact solutions \cite{guo2021exact,sayyad2023topological} 
or the single-particle picture \cite{mu2020emergent,alsallom2022fate,kawabata2023entanglement} to numerically obtain exact many-body ground states and dynamics. 
It is well-known that a non-Hermitian Hamiltonian violates the conditions of quantum mechanics, for which
the $\mathcal{PT}$ symmetric quantum mechanics were introduced to ensure the real eigenvalues and the unitarity \cite{bender1998real,bender2002complex}.
However, recent studies indicate a non-Hermitian system with complex eigenvalues can in principle exhibit richer physics \cite{bergholtz2021exceptional,ashida2021non}.
Therefore, it is worth while to understand the construction of many-body spectra and low-energy excitations of a non-Hermitian many-body system from its complex single-particle spectrum.

\begin{figure}[t]
\centering
\includegraphics[width=8.6cm]{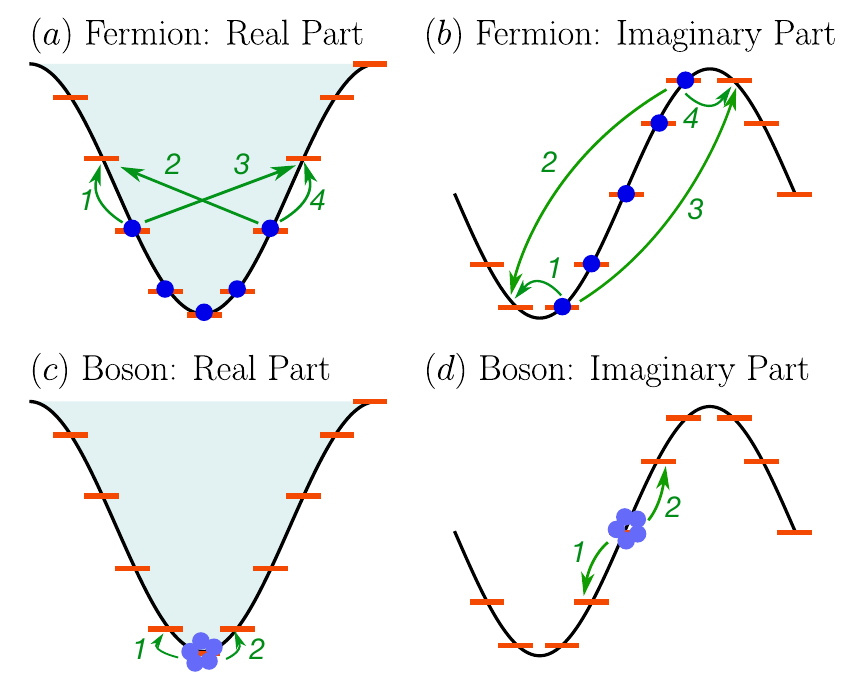} 
\caption{
Configurations of particles of many-body ground states and low excited states of the HN model in PBC for $L=10$ lattice sites and $N=5$ particles.
(a),(b) The real and imaginary parts of the ground state and four-fold degenerate first excited states of fermions.
(c),(d) The real and imaginary parts of the ground state and two-fold degenerate first excited states of bosons.
}
\label{Fig:aufbau}
\end{figure}

On the other hand, the precision of measurements is governed by the Heisenberg uncertainty principle \cite{giovannetti2004quantum}, which states that one cannot exactly measure the position and the momentum of a particle at the same time.
For instance, when ideal bosons condense at the momentum space, the position distribution of bosons must be extended. 
A strategy to increase the accuracy is to design a quantum state where the uncertainty of one observable is very small \cite{giovannetti2004quantum}.
It is known that when a non-Hermitian system exhibits the non-Hermitian skin effect, all of the single-particle eigenstates can be exponentially localized at the boundary.
An interesting question is that the momentum distribution of particles should be extended or localized if the system is prepared in a localized state through the non-Hermitian skin effect?

In this paper, we address these questions by employing the generalized Aufbau principle introduced below, from which full eigenvalues and many-body states of a system can be exactly constructed from single-particle energy levels.
The generalized Aufbau rule is demonstrated to be identical to that developed in Hermitian systems when the complex single-particle energies are labelled according to their real parts.
Furthermore, we find that non-Hermitian many-body skin effects can persist in all of the many-body eigenstates in both fermionic and bosonic systems, 
where the momentum and position distributions are argued to be independent unexpectedly.
That is to say bosons can simultaneously localize in both the real and momentum space, revealing an anomalous Bose-Einstein condensation beyond the Heisenberg uncertainty principle. 
Finally, we show that bosons under the infinite interaction (in the hard-core limit) cannot be mapped to fermions in periodic boundary conditions.

{\it Generalized Aufbau principle.-} 
The Aufbau principle that illustrates the rules how particles full up orbitals according to the laws of quantum mechanics play an important role in solid-state physics, statistical physics and quantum chemistry.
The achievement of the Aufbau principle depends largely on the Hermiticity of quantum mechanics, where the real lowest and higher-energy levels can be efficiently labelled. 
However, in non-Hermitian systems the spectra are complex, which seems trouble to label or even meaningless to fill up these complex-valued energy levels.
In the following, we will show that how to construct the configurations of indistinguishable particles and find interesting non-Hermitian physics based on the generalized Aufbau principle.
Unexpectedly, the generalized Aufbau principle is found to be identical to that developed in Hermitian systems. 
That is to say one can investigate rich unique phenomena of non-Hermitian many-body systems by using the well-established approaches developed in Hermitian systems
if the nonorthogonality from the non-Hermiticity is correctly considered.

Let us now introduce the generalized Aufbau principle. Given an arbitrary non-Hermitian canonical system composed of $L$ energy levels and $N$ spinless particles,
the single-particle energy eigenvalues of a system are assumed as $\varepsilon_{m}=\varepsilon_{m}^{R} + i \varepsilon_{m}^{I}$, where $\varepsilon_{m}^{R}$ and $\varepsilon_{m}^{I}$
are the real and imaginary parts of $\varepsilon_{m}$ with $m=1, \cdots, L$. 
The Aufbau principle for non-Hermitian systems, named as \textit{generalized Aufbau principle}, states that particles first fill the level of the lowest real part of energy, 
then fill the levels of the higher real parts of energy governed by quantum statistics independent on the imaginary parts of energy. 
In the following, we employ this generalized Aufbau principle in the Hatano-Nelson model to illustrate how to construct full many-body spectra based on the single-particle spectrum.

{\it Model.-} The non-Hermitian system we considered as an example is the nonreciprocal Hatano-Nelson (HN) model \cite{hatano1996localization,hatano1997vortex} in one dimension given by,
\begin{align}
H_{\text{HN}} = t \sum_{j=1}^{L} \left( e^{g} c_{j}^{\dagger}c_{j+1} + e^{-g} c_{j+1}^{\dagger}c_{j} \right),
\label{eq:HNmodel}
\end{align}
where $L$ is the length of the chain, $c_{j}^{\dagger}$ and $c_{j}$ are the creation and annihilation operators of spinless fermions or bosons at the $j$th lattice site,
$t>0$ and $g>0$ are two nearest neighbor coupling constants.
The HN model in Eq.(\ref{eq:HNmodel}) is a non-Hermitian tight-binding model as $H_{\text{HN}} \neq H_{\text{HN}}^{\dagger}$. 
The periodic boundary conditions (PBC) of the chain is imposed by the condition $c_{L+1}=c_{1}$.

The single-particle energy spectrum of the HN model for the finite chain in PBC can be obtained as,
\begin{align}
\varepsilon_{k_{m}} = te^{g}e^{-ik_m} + te^{-g}e^{ik_m},
\label{eq:spectrumPBC}
\end{align}
by using the Fourier transformation $c_{j}=\frac{1}{L} \sum\limits_{k_m}e^{-ik_{m}j}c_{k_m}$. The wave number $k_m=2\pi m/L$ with $m=1, \cdots, L$.
The eigenenergies in Eq.(\ref{eq:spectrumPBC}) are complex except $m=L/2, L$ ($m=L$) for even (odd) lattice sites.
The real and imaginary parts of the energy eigenvalues are the cosine function $\varepsilon_{k_{m}}^{R} = (e^{g}+e^{-g}) \cos k_m$ 
and the sine function $\varepsilon_{k_{m}}^{I} = (-e^{g}+e^{-g}) \sin k_m$ as shown in Fig.\ref{Fig:aufbau}.

\begin{figure}[t]
\centering
\includegraphics[width=8.9cm]{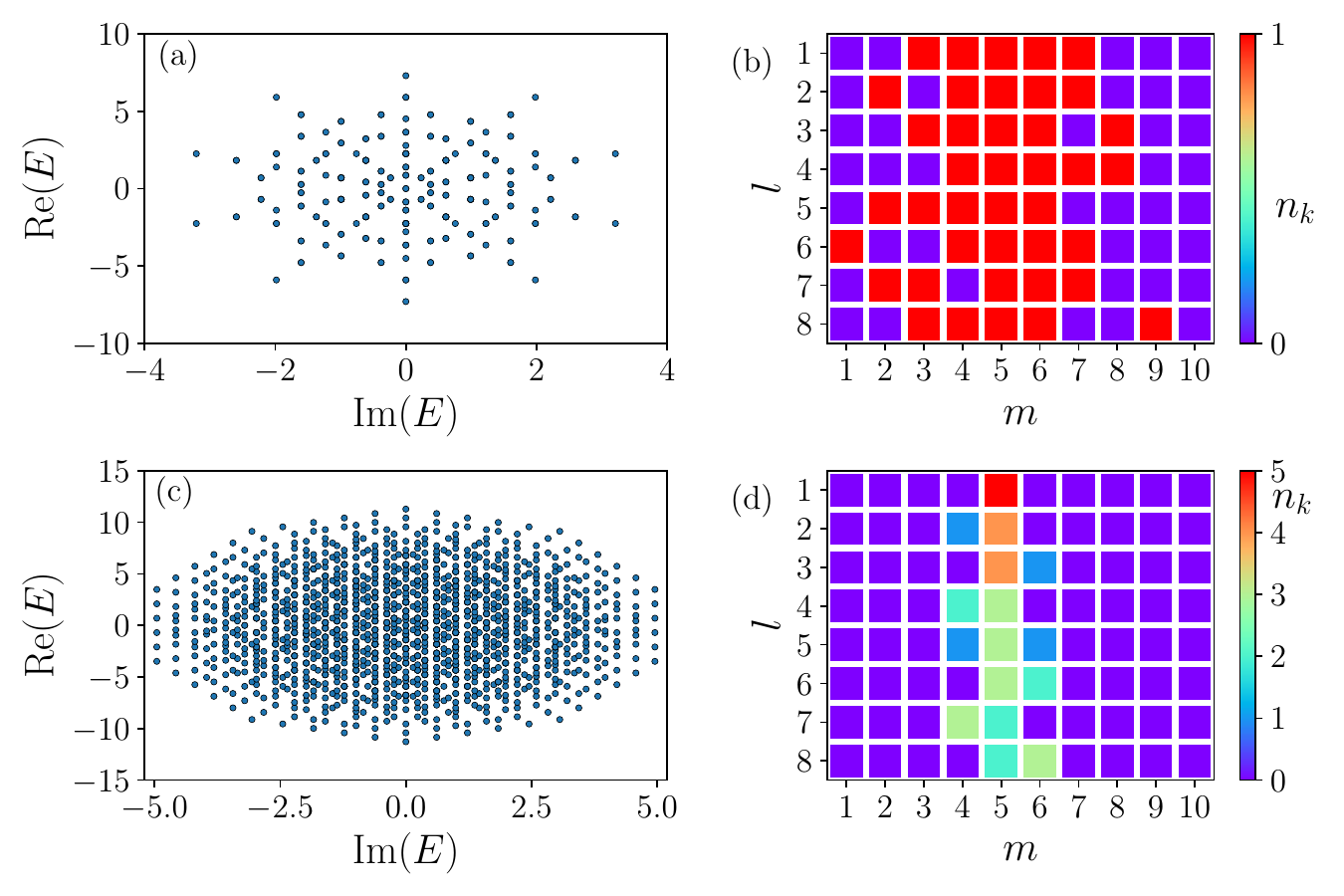} 
\caption{
Many-body energy spectra and momentum distributions of the fermionic and bosonic HN models in PBC for $L=10$ lattice sites and $N=5$ particles with $g=0.5$, $t=1$.
(a) The full many-body energy spectrum for fermions.
(b) The momentum distributions $n_k$ of the fermionic HN model for lowest eight eigenstates.
(c) The full many-body energy spectrum for bosons.
(d) The momentum distributions $n_k$ of the bosonic HN model for lowest eight eigenstates.
}
\label{Fig:HNspetrum}
\end{figure}

{\it Many-body spectra.-} 
Let us start with a many-body system with $N$ particles and $L$ lattice sites described by Eq.(\ref{eq:HNmodel}) in PBC. 
As can be seen from Eq.(\ref{eq:HNmodel}), particle numbers $N=\sum\limits_{j}c_{j}^{\dagger}c_{j} $ in the system 
are conserved because $\left[N, H_{\text{HN}} \right ]=0$, although the Hamiltonian is non-Hermitian. 
As the system in Eq.(\ref{eq:HNmodel}) has $L$ single-particle states, the total number of particles and the total energy of a many-body state is,
\begin{align}
N &= \sum_{m=1}^{L} n_{k_{m}}, \\
E &= \sum_{m=1}^{L} n_{k_{m}} \varepsilon_{k_{m}},
\end{align}
where the particle numbers $n_{k_{m}}=0,1$ for fermions and $n_{k_{m}}=0,1, \cdots, N$ for bosons in each energy level $\varepsilon_{k_{m}}$.

Assuming particles can occupy states with complex-valued eigenenergies,
the total size of the $N$-particle Hilbert space (or the total number of many-body eigenenergies $E_i$) for a non-Hermitian system with $N$ particles in $L$ states will be $\frac{L!}{N!(L-N)!}$ and $\frac{(L+N-1)!}{N!(L-1)!}$ for fermions and bosons as Hermitian systems, respectively.
To reveal the physics of these many-body states, we sort the real parts of single-particle energies in ascending order given as,
\begin{align}
\varepsilon_{p_{1}}^{R} \leq \varepsilon_{p_{2}}^{R} \leq \cdots \leq \varepsilon_{p_{L}}^{R},
\label{eq:RealPartPBC}
\end{align}
where $p_{l} \in \{k_{m}\}$. 
It should be noted that the reason for using real parts to construct the many-body energy spectrum is that (i) filling single-particle states with purely imaginary energies is unphysical \cite{chang2020entanglement}, 
and (ii) the longest-surviving state of one Hamiltonian with the largest imaginary part of the energy can be mapped the ground state of another Hamiltonian with the lowest real part of the energy \cite{yamamoto2022universal,banerjee2022chiral}.

As can be seen in Fig.\ref{Fig:aufbau}, the ground state (the lowest real part of the energy $E_0$) of the fermionic HN model is regarded as such a state 
in which particles fill up the first $N$ lowest single-particle state $\varepsilon_{p_l}$ with $l=1,2,3,4$.
While for bosons, the ground state is that state where all of the particles condense only at the single-particle state $\varepsilon_{p_1}$ with the lowest real part of the energy.
The excitations can be obtained by moving particles to the levels with higher real parts of energies as demonstrated in Fig.\ref{Fig:aufbau} and Fig.\ref{Fig:HNspetrum}, 
in which four (two) first excited states for fermions (bosons) are found in an even lattice $L$ with an odd $N$ as expected.
There is nothing surprising in this analysis as identical particles should obey quantum mechanics.
Consequently, it seems that interesting non-Hermitian physics are hidden in this approach as one cannot distinguish constructions between the Hermitian and the non-Hermitian systems.

\begin{figure}[t]
\centering
\includegraphics[width=8.6cm]{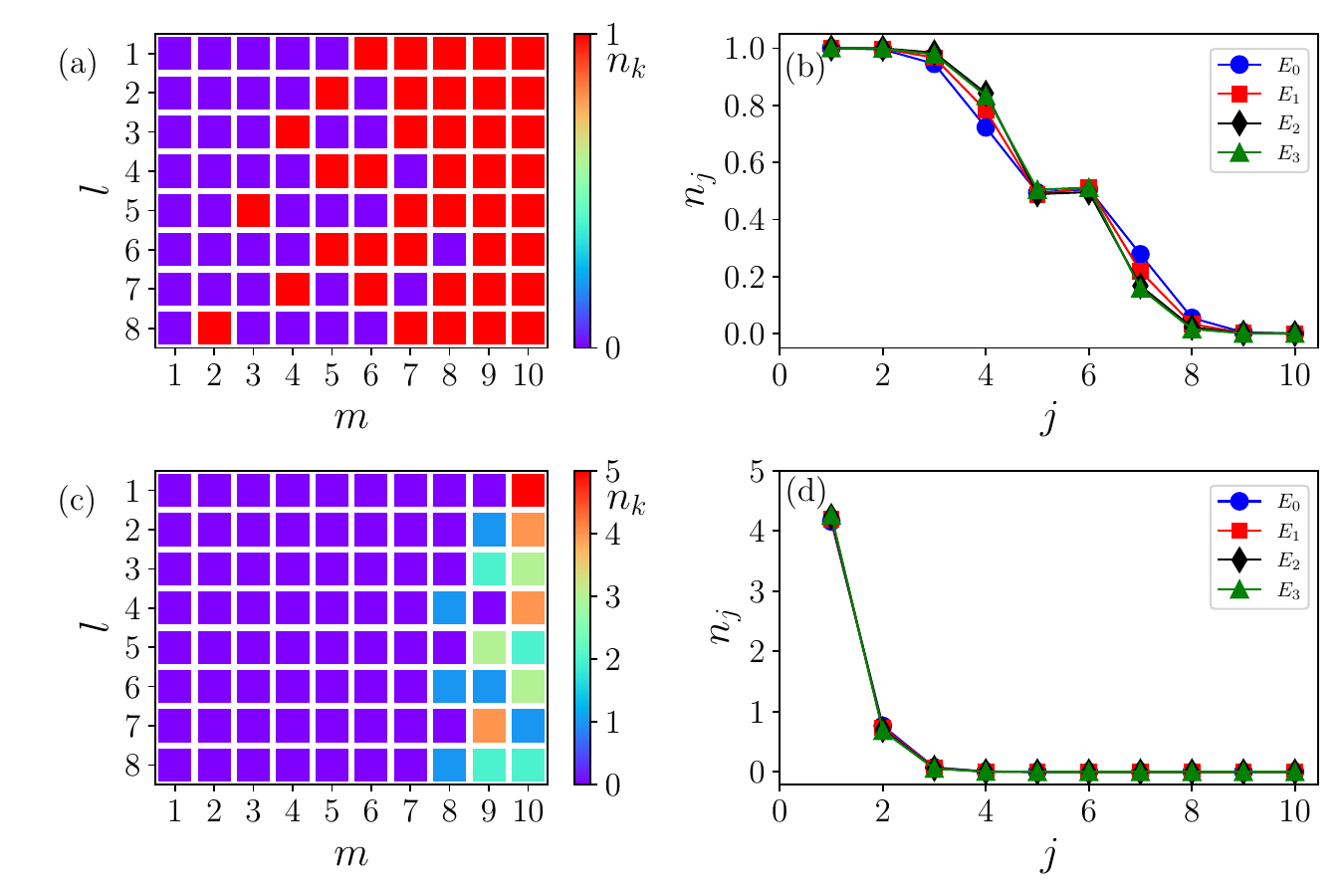} 
\caption{
Position and momentum distributions of the fermionic and bosonic HN models in OBC for $L=10$ lattice sites and $N=5$ particles at $g=1.5$, $t=1$.
(a) The momentum distributions $n_k$ of the fermionic HN model for lowest eight eigenstates.
(b) The position distributions $n_j$ of the fermionic HN model for lowest four eigenstates.
(c) The momentum distributions $n_k$ of the bosonic HN model for lowest eight eigenstates.
(d) The position distributions $n_j$ of the bosonic HN model for lowest four eigenstates.
}
\label{Fig:HNskin}
\end{figure}

{\it Many-body skin effect.-} 
In order to illustrate the unique physics in non-Hermitian systems, we turn to the open boundary condition (OBC). There is no doubt that above analysis can apply to a system in OBC. 
The single-particle spectrum in OBC can be simply derived as \cite{yang2022designing}, 
\begin{align}
\epsilon_{k_{m}}=2t\cos k_{m}^{\prime},
\label{eq:spectrumOBC}
\end{align}
by replacing the $k_m \rightarrow k_{m}^{\prime} -i g$ according to the non-Bloch band theory \cite{yao2018edge,yokomizo2019non}, where $k_{m}^{\prime} = m \pi /(L+1)$ and $m=1, \cdots, L$.
It is evident that the eigenvalues of Eq.(\ref{eq:spectrumOBC}) are always real as the Hamiltonian in Eq.({\ref{eq:HNmodel}}) 
can be mapped to a Hermitian Hamiltonian under a site-dependent similarity transform $c_{j}^{\prime}=c_{j}e^{-jg}$ in OBC.
The configurations of particles for the first eight eigenstates are demonstrated in Fig.\ref{Fig:HNskin} (a) \& (c), where we find similar structures shown in PBC.

To reveal the unique phenomenon of non-Hermitian systems, such as the non-Hermitian skin effect, we turn to analyzing many-body eigenstates of the HN model in OBC.
Let $\ket{\psi} =(\phi_1, \phi_2, \cdots, \phi_L)^\text{T}$ denote single-particle states, which satisfy the real-space eigenvalue equation $H \ket{\psi} = E \ket{\psi}$.
The single-particle eigenstates are given as \cite{yokomizo2019non,mu2020emergent,guo2021exact},
\begin{align}
\ket{\psi_{m}} =(\phi_{1}^{m}, \phi_{2}^{m}, \cdots, \phi_{L}^{m})^\text{T}
\end{align}
with $\phi_{j}^{m}=r^j \sin(j k_{m}^{\prime})$ and $r=e^{-g}$.
Suppose there are $N$ particles $q_{i}$ occupying $N$ single-particle states $\ket{\psi_{l}}$, with $i, l=1,2, \cdots, N$, the many-body state for fermions is, 
\begin{align}
\ket{\Psi^{f}} = A_{f} \sum_{p}  \text{sgn}(p) \ket{\psi_{1}(q_1)} \ket{\psi_{2}(q_2)} \cdots \ket{\psi_{N}(q_N)},
\label{Eq:WaveFermion}
\end{align}
where $p$ denotes the permutations acting on $N$ particles, $A_{f}$ denote the coefficients of normalization, 
$\text{sgn}(p)$ is the sign function from the permutations due to the fermionic anti-commutation relations.
If the particles are bosons, the many-body state is a totally symmetric state, which is given by,
\begin{align}
\ket{\Psi^{b}} = A_{b} \sum_{p} \ket{\psi_{1}(q_1)} \ket{\psi_{2}(q_2)} \cdots \ket{\psi_{N}(q_N)}.
\label{Eq:WaveBoson}
\end{align}
It should be noted that the coefficients $A_{f}$ and $A_{b}$ are not the same as those in Hermitian systems due to the non-orthonormality of the eigenstates $\ket{\psi_{l}}$.
However, the coefficients $A_{f}$ and $A_{b}$ that can be numerically computed for arbitrary tight-binding models play no role in understanding the unique phenomena of non-Hermitian systems
as observables should be computed under the normalization. 
Interesting non-Hermitian physics of a Hamiltonian originate from the nonorthogonality of the $\phi_{j}^{m}$ \cite{kawabata2023entanglement}.

To illustrate the localization of particles, we calculate the position distributions,
\begin{align}
n_j = \bra{\Psi_{f,b}} \hat{n}_{j} \ket{\Psi_{f,b}},
\end{align}
for each lattice site obtained from Eq.(\ref{Eq:WaveFermion}) and Eq.(\ref{Eq:WaveBoson}), as shown in Fig.\ref{Fig:HNskin} (b) \& (d).
We find that the fermionic many-body states cannot exponentially localize at the boundary but exhibit an asymmetry due to the Pauli exclusion principle. 
This phenomenon is the many-body version of the non-Hermitian skin effect \cite{alsallom2022fate,zhang2022symmetry}. 
In contrast, all of the bosons can exhibit exponential localization at the boundary for arbitrary many-body states independent on the momentum distributions $n_k$ [c.f. Fig.\ref{Fig:HNskin} (c) \& (d).].
That is to say bosons can simultaneously localize in both the real and momentum space, which is confirmed in the ground state of the bosonic HN model [c.f. Fig.\ref{Fig:HNskin} (c) \& (d)].
This unique feature of non-Hermitian skin effect, named as anomalous Bose-Einstein condensation, can explain why the skin superfluid may appear \cite{zhang2020skin}, 
and may potentially help to improve the precision of measurement in quantum metrology, which is left for future study.

\begin{figure}[t]
\centering
\includegraphics[width=8.8cm]{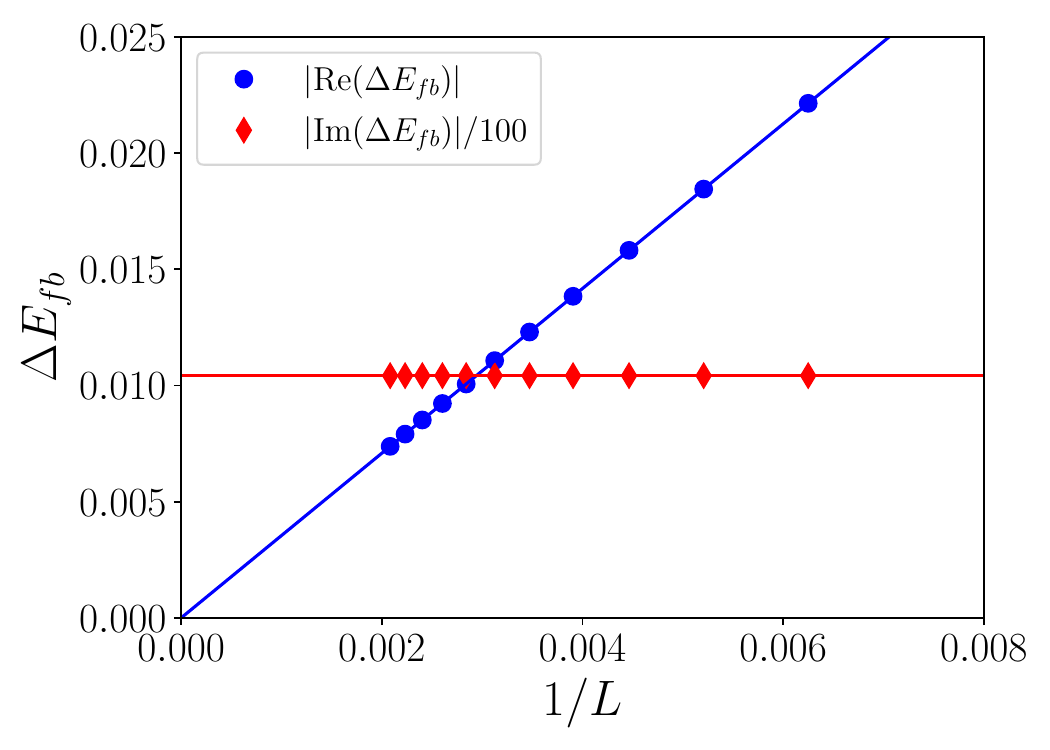} 
\caption{
The differences of ground-state energies between fermions and hard-core bosons of the HN model in PBC at half filling at $g=0.5$, $t=1$. The system ranges in size from $L=160$ to $L=480$.
The blue circles and red diamonds denote the real parts and imaginary parts of the energy difference, respectively. A factor $1/100$ is used for imaginary parts of the energies for comparison.
}
\label{Fig:FBcompare}
\end{figure}

{\it Fermions and hard-core bosons.-} 
Let us finally discuss interacting bosons in the hard-core constraint in a chain that can be realized by imposing an infinite repulsive on-site interaction. 
For simplicity, we consider a system with an even lattice $(L=2K)$ in half filling $(N=K)$, where $K$ is positive integer.
It is well-known that hard-core bosons can be mapped to noninteracting spiness fermions by the Jordan-Wigner transformation \cite{nie2013ground} under OBC in Hermitian tight-binding models.
For PBC, hard-core bosons are equivalent to spinless fermions if systems have an odd integer number of particles ($K$ is an odd integer).
However, hard-core bosons are not equivalent to fermions for the system with an even number of particles ($K$ is an even integer), 
in which the hard-core bosons in PBC are mapped to spinless fermions with the antiperiodic boundary condition \cite{nie2013ground}. 
Moreover, hard-core bosons in PBC have lower energy with the order $O(1/L^2)$ than spinless fermions in PBC \cite{nie2013ground}, 
indicating that the hard-core boson is equivalent to the spinless fermions in the thermodynamic limit.

For the HN model, we find that the equivalence between hard-core bosons and spinless fermions remains in both OBC and PBC, which is not surprising as bosons and fermions should obey same statistics as that in Hermitian systems.
It seems that the hard-core bosons should be equivalent to the spinless fermions in one-dimensional non-Hermitian systems in the thermodynamic limit.
However, as can be seen from Fig.\ref{Fig:aufbau}, if we consider a finite chain in PBC with an even number of particles $N$, 
the many-body ground state of the fermionic HN model should be doubly degenerate with complex-valued energies.
While for bosons, the many-body ground state is always real even in the hard-core limit due to the statistics of bosons.
This implies that hard-core bosons in an even lattice $L=2N$ with even particles $N$ cannot be mapped to spinless fermions in PBC.

To verify whether such non-equivalence remains in the thermodynamic limit, we compute the differences of ground-state energies,
\begin{align}
\Delta E_{fb} = E_{0}^{f} - E_{0}^{b},
\end{align}
between fermions and hard-core bosons in half filling for the HN model up to $L=480$ with an even number of particles $N$ under PBC, 
where $E_{0}^{f}$ and $E_{0}^{b}$ are the ground-state energies of fermions and hard-core bosons.
We find that the real parts of $\Delta E_{fb}$ decease linearly with the chain length as demonstrated in Fig.\ref{Fig:FBcompare} as expected in Hermitian systems.
While we unexpectedly find that the imaginary parts of $\Delta E_{fb}$ remain constant over the increase of the chain length.
The exact value of the imaginary part of the energy difference $\Delta E_{fb}$ is found as,
\begin{align}
\text{Im}(\Delta E_{fb}) = (-e^{g} + e^{-g}) \sin \pi \left( 1-\frac{N}{L} \right),
\end{align}
dependent only on the filling $N/L$.
This indicates that hard-core bosons in PBC in an even lattice $L$ consisting of even particles $N$ are not equivalent to spinless fermions in PBC even in the thermodynamic limit.
We therefore note that one should take care to perform the Jordan-Wigner transformation to map the fermions to spins as non-Hermitian systems can be very sensitive to boundary conditions \cite{guo2021exact}.

{\it Conclusion.-} 
In summary, we have developed the generalized Aufbau principle to study the many-body physics of non-Hermitian systems with complex-valued energy eigenvalues.
It is shown that the generalized Aufbau principle can be used to obtain all of the non-Hermitian many-body phenomena if the nonorthogonality of the single-particle eigenstates is correctly considered.
Many-body skin effects that are numerically confirmed are argued to appear for both the bosons and the fermions in all of the many-body eigenstates. 
In particular, it is shown that the ground state of bosons is an anomalous Bose-Einstein condensation with particles simultaneously localizing in both the real and momentum space.
Furthermore, we find that the fermions and the hard-core bosons are different in PBC. 
In the future, more non-Hermitian models would be explored by using the generalized Aufbau principle and other well-established methods developed in Hermitian quantum theory.

{\it Acknowledgments.-} 
G. S. was supported by the NSFC under the Grant No. 11704186 and "the Fundamental Research Funds for the Central Universities, NO. NS2023055".
S.-P. K is appreciative of support from the NSFC under the Grant Nos. 11974053 and 12174030.
This work is partially supported by High Performance Computing Platform of Nanjing University of Aeronautics and Astronautics.


\bibliography{ref}

\end{document}